\documentclass[
 reprint,
 amsmath,amssymb,
 aps,twocolumn,
pra
]{revtex4-1}

\usepackage{dcolumn}
\usepackage{amsbsy}
\usepackage{amsmath}

\usepackage{amssymb}
\usepackage{braket}
\usepackage{color}
\usepackage{fullpage}
\usepackage{graphicx}
\usepackage{bm}

\usepackage{subfigure}
\usepackage{hyperref}
\usepackage{cleveref}

\usepackage{epstopdf}

\begin{document}

\title{Temporal correlation beyond quantum bounds in non-Hermitian dynamics }

\author{Anant V. Varma}%
 \email{anantvijay.cct@gmail.com}
 
\author{Ipsika Mohanty}%
\email{ipsika.mohanty@gmail.com}

\author{Sourin Das}
 \email{sourin@iiserkol.ac.in , sdas.du@gmail.com}

\affiliation{%
Indian Institute of Science Education \& Research Kolkata,
Mohanpur, Nadia - 741 246, 
West Bengal, India
% IISER-Kolkata, Mohanpur,741246, West Bengal, India.
}%

\begin{abstract}
We study the dynamics of a non-Hermitian Hamiltonians two level system  (TLS) with real eigenvalues. Within the framework of Hermitian quantum mechanics, it is known that maximal violation of Leggett-Garg inequality (LGI) is bounded by $K_{3}=3/2$ (Luder's bound). We show that this absolute bound can be evaded when dynamics is governed by non-Hermitian Hamiltonians with real eigenvalues. Moreover, the extent of violation can be optimized to asymptotically approach the algebraic maximum of $K_{3}^{max}=3$, which is otherwise being observed for Hermitian Hamiltonian with infinite dimensional Hilbert space. The extreme violation of LGI is shown to be directly related to the two basic ingredients: {\it {(i)}} The Bloch equation for the TLS having a non-linear terms which allow for accelerated dynamics of states on the Bloch sphere exceeding all known quantum speed limits of state evolution; and {\it{(ii)}} Quantum trajectory of states lies on a great circle (geodesic path) on the Bloch sphere at all times. Finally, we demonstrate that such extreme temporal correlation of TLS can be simulated in realistic system by embedding the TLS into a higher dimensional Hilbert space such that the composite system obeys unitary dynamics. Specifically we show that a four dimensional embedding of non-Hermitian TLS is enough to host $K_{3} \rightarrow 3$ limit.  These findings suggest that a qubit coupled to external quantum degrees of freedom with prescribed interaction between them, i.e., an appropriate embedding, can serve as an ideal resource for speeding up of quantum operations on the qubit and hence can result in unprecedented speed ups in quantum information processing times.
%These findings indicate that an appropriate finite dimensional embedding of a qubit (TLS) into an higher dimensional Hilbert space could serve as a resource for speed up of quantum gate operations on qubit and hence resulting in faster quantum information processing.}
\end{abstract}
%\keywords{Suggested keywords}%Use showkeys class option if keyword
                              %display desired
\maketitle

%\tableofcontents

\section{\label{sec:level1}Introduction}
%%%%%%%%%%%%%%%%%%%%
The subatomic world of particles follows the laws of quantum mechanics, while at macroscopic scales much of the phenomena are described by classical mechanics. There has been a persistent quest to understand the validity of quantum mechanics at the macroscopic scales. LGI has emerged as temporal analog of Bell's inequality~\cite{Bell1964,Peres1999,Mahler1993,Budroni2013,Kim2006} which provides test of quantum mechanics at macroscopic scales via violation of bounds on temporal correlations~\cite{LG1985,Leggett2008,Legget2002}. Macroscopic realism (MR) and non-invasive measurability (NIM) underlies the basic construction of LGI~\cite{LGIreview2014},  which are based on our intuition of the classical world and are not conformed by quantum mechanics. Thus, violation of the LGI surely indicates a breakdown of any one of the above assumptions or both. And, hence its violation can be considered as an indication  of non-classical or quantum behaviour under appropriate experimental circumstances~\cite{Mahler1993,Wilde2012}. Currently, there are theoretical studies~\cite{Kunkun2018,Aditi2018,Zhan2017,MAL2016,Kofler2013,Suzuki2012,Pan2018,Naikoo2019} as well as experimental confirmation of violation of LGI
 in a large variety of macroscopic systems ranging from superconducting qubits~\cite{Palacios2010,Xiang2013} to nuclear spins~\cite{Athalye2011,Formaggio2016,Katiyar2013}.\\
Explicit construction of LGI for multi-level quantum system involves identification of a dichotomic observable $Q$ such that its eigenvalues are restricted to $\pm 1$. LGI denoted by $K_{3}$ with the assumption of MR and NIM is bounded as  $-3 \leq K_{3}=C_{12}+C_{23}-C_{13}\leq 1$. Here $ C_{ij}=\sum_{Q(t_i),Q(t_j)=\pm 1} Q(t_i) Q(t_j)P_{ij}(Q(t_i),Q(t_j))$, where $Q(t_i),Q(t_j)$ represents the outcomes of the strong quantum measurements of the observable $Q$ at times $t_i$ and $t_j$ respectively and $P_{ij}(Q(t_i),Q(t_j))$ represents the joint probability for the outcome of quantum measurement performed at time $t_i$ and  $t_j$ to be $Q(t_i)$, $Q(t_j)$ respectively. The maximum quantum bound of $K_3$ for any $N$ level system is $3/2$ which is known as the Luder's bound ~\cite{Budroni2013}. Violation of Luder's bound for an $N$  level quantum system, where $N>2$ is possible provided further degeneracy breaking measurements are introduced \cite{Emary2014} but violation of Luder's bound for $N=2$ i.e. TLS is impossible within the  Hermitian dynamics. \\
In this article, we demonstrate the possibility of violation of the Luder's bound of $3/2$ and asymptotically approaching the algebraic bound of $3$ for the  LGI parameter ($K_{3}$) for a TLS.  We show that the dynamics of a TLS governed by a non-Hermitian Hamiltonian with real eigenvalues ~\cite{HFJones2002,Bender1998} leads to such extreme violation of LGI. The implications of our findings are non-trivial in two ways: {\it(a)}  violation of the Luder's bound provides a clear quantification for temporal correlation which are more correlated then what is allowed by Hermitian dynamics of TLS  {\it(b)} approaching the maximum value of $K_3=3$ for LGI is a clear indicator of extreme temporal correlation within the Hilbert space of TLS, which has only been observed in the past for Hermitian quantum mechanics for Hilbert spaces dimensions tending to infinity~\cite{Emary2014}.\\
%%%%%%%%%%%%%%%%%%%%%%%%%%%%%%%%%%%%%%
\subsection{Probabilistic considerations of $K_3=3$ } %%%%%%%
%%%%%%%%%%%%%%%%%%%%%%%%%%%%%%%%%%%%%%
We start by noting that the maximum violation of LGI corresponding to $K_3=3$ imposes stringent constraints on the joint probabilities $P_{ij}(Q(t_i),Q(t_j))$ which can not be satisfied within pure Hermitian dynamics of TLS. This can be understood as follows. Measurement of LGI involves quantum measurement at three different times (call it $t_1,t_2,t_3$ such that $t_1<t_2<t_3$) with successive measurement performed in pairs.  Now, $K_3=3$ implies that $C_{12}=C_{23}=1$ while  $C_{13}=-1$.  $C_{ij}=1$  in turn implies that joint probabilities $P_{ij}(+,+) + P_{ij}(-,-) =1$ and $P_{ij}(+,-) = P_{ij}(-,+) =0$, owing to the  normalization constraints satisfied by the probability given by $\sum_{Q(t_i),Q(t_j)=\pm 1} P_{ij}(Q(t_i),Q(t_j))=1$.  Similarly, $C_{ij}=-1$ implies that   $P_{ij}(+,+) = P_{ij}(-,-) =0$ and  $P_{ij}(+,-) + P_{ij}(-,+) =1$. Hence the condition of $C_{ij}=-1$ and $C_{ij}=1$ are mutually exclusive in the sense that the former condition needs $P_{ij}(+,-) = P_{ij}(-,+) =0$ and rest to be finite while the later condition demands just the opposite. And hence the condition for $K_3=3$ implies that the joint probability of observation of ``spin flip state" between times $t_1, t_2$ given by $P_{12}(-,+)/P_{12}(+,-)$ and between the times $t_2, t_3$  given by $P_{23}(-,+)/P_{23}(+,-)$ have to be identically zero while the joint probability of observation of "same spin state" between time $t_1, t_3$ given by $P_{13}(+,+)/P_{13}(-,-)$ have to identically zero. \\
Now, let us consider an ideal situation where the three inputs, the initial state, the Hamiltonian and the times $t_1,t_2,t_3$  are chosen in such a way that the time evolution of the initial state, when not interrupted by any measurement, evolves into eigenstate of the dichotomic observable $Q$ at  $t_1,t_2,t_3$. Even under this ideal situation where the measurement is non-invasive in the sense that measured state is ensured to be an eigenstate of measurement operator itself, satisfying the condition of $C_{12}=C_{23}=1$ and  $C_{13}=-1$ simultaneously is impossible as no physically realizable dynamics can produce states which can simultaneously  satisfy $P_{12}(-,+)/P_{12}(+,-)= P_{23}(-,+)/P_{23}(+,-)=0$ and $P_{13}(+,+)/P_{13}(-,-)=0$. So the interesting question that one can ask  is if there is any possibility for approaching asymptotically closeness to this condition. Again, based on above argument it is straightforward to see that Hermitian dynamics of a TLS can not do this because the speed of evolution of non-eigenstate of TLS on the Bloch sphere is always uniform (constant in time). The crucial element which is essential for approaching $K_3=3$ is a nonuniform, accelerated (or decelerated) evolution of state between two successive quantum measurements. \\
In what follows, we first discuss the equation of motion (the Bloch equation) for  pseudo spin corresponding to TLS evolving under the influence of non-Hermitian Hamiltonian with real eigenvalues. We identify the presence of non-linear term in Bloch equation which is responsible for such accelerated evolution and obtain analytic solution of the equation.  For identifying the parameter space of $K_3=3$,  using these solutions we first identify the subspace of parameters corresponding to $C_{13}=-1$ and then look for the point in the subspace which also satisfies $C_{12}=C_{23}=1$.   \\
%%%%%%%%%%%%%%%%%%%%%%%%%%%%%%%

\section{\label{sec:level2 }Non-Hermitian Bloch equation } %%%%%%
%%%%%%%%%%%%%%%%%%%%%%%%%%%%%%%
A general non-Hermitian Hamiltonian for a TLS can be written as $H = (\vec{A} - i \vec{B}) \cdot \vec \sigma $,  where $\vec\sigma$ represents a vector whose components are given by Pauli matrices and $\vec{A} , \vec{B} $ are vectors which live in ${\mathbb{R}}^3$ and to ensure that the eigenvalues of $H$ stay real we impose that $\vec{A} \cdot \vec{B}=0$ and $\vert \vec{A}  \vert > \vert \vec{B} \vert$  ~\cite{Jordan1969}. Such non-Hermitian systems have been shown to exhibit non trivial effects such as acceleration in the speed of the evolution of a quantum state  \cite{Jones2007}. 
%, recovery of entanglement \cite{SLChen2014} and state discrimination \cite{Royal2013}. 
 If any arbitrary initial state is evolved through such non-Hermitian Hamiltonian, then the governing equation is given by \cite{Graefe2012}  :

\begin{equation}
    \frac{d \rho_{t}}{dt} =  -i \left [ \vec{A}\cdot \vec \sigma,\rho_{t} \right ] -\left \{ \vec{B}\cdot \vec \sigma,\rho_{t}  \right \} + 2 \ tr( \rho_{t}  \vec{B}\cdot \vec \sigma) \rho_{t}
    \label{0}
\end{equation}

It is interesting to note that this type of time evolution as in eq. (\ref{0}) has been extensively discussed in many contexts like constrained quantum motion \cite{Brody1}, approach to thermal equilibrium \cite{Korsch}, non-Hermitian quantum motion and  dissipation \cite{Mizrahi,Sergi}, and also has been realized experimentally \cite{Tang,Li,Peng}. The formal solution of the eq.(\ref{0}) is given by:

\begin{equation}
    \rho_{t} = \rho_{t} = (1/2)~  \mathbb{I} + \vec{S}(t) \cdot \vec \sigma = \frac{e^{- i H t} \rho_{0} e^{i H^{\dagger} t}}{tr (e^{- i H t} \rho_{0} e^{i H^{\dagger} t})} 
    \label{1}
\end{equation}

An experimental implementation of such a evolution requires a careful balancing of
loss and gain of energy of the quantum system such that eigenvalues of the Hamiltonian remain real at all times. It should be evident, however, that the physical realisation of such an evolution is difficult since it requires extreme fine-tuning. The extension of eq.(\ref{0}) which takes account of the possible impact of uncontrollable ambient noise in energy has been studied in \cite{Graefe2012}, and we will also implement it while discussing the possible effects of such noise on the violation of Luder's bound in later section.\\

The Bloch equation dictating the time evolution of the vector $\vec{S}$ corresponding to eq.(\ref{0}) is given by: 

%%%%%%%%%%%%%%%%%%%%%%%%%%%%%%%%%%%%%%%%%%%%%%%%%%%
\begin{equation}
\frac{d\vec{S}(t)}{dt}  = 2 \ \vec{A}\times \vec{S}(t)  -  \vec{B} + 4 \  {\{} \vec{B} \cdot \vec{S}(t)  {\}} \ \vec{S}(t)~.
\label{2}
\end{equation}

%%%%%%%%%%%%%%%%%%%%%%%%%%%%%%%%%%%%%%%%%%%%%%%%%%%
Time evolution in eq.(\ref{2}) is such that it preserves the length of the vector $\vec{S}(t)$ in case of pure states only. Note the appearance of a term which is non-linear in $\vec{S}$ in eq.(\ref{2}). This is an unusual addition to the standard Bloch equation which usually includes the precession term (first term in eq.(\ref{2})) and the decay type term (second term in eq.(\ref{2})). The third term  in eq.(\ref{2}) is the one which is responsible for an accelerated (or decelerated) evolution of  pure states on the Bloch sphere which is a crucial ingredient for maximizing the $K_3$ as described above. Further we would like to emphasize that this equation has solution which forms periodic orbits on the Bloch sphere with periods determined by the difference in eigenvalues of $H$  given by $\pm\sqrt{\vec{A}^2 - \vec{B}^2}$ just like  the Hermitian case. To obtain analytical solutions of eq.~\ref{2}, we work in a Cartesian coordinate system defined by the unit vectors $\hat{A}$, $\hat{B}$ and $\hat{n}= \hat{A}\times \hat{B}$. Rewriting eq.(\ref{2}) in its component form, we get 
%%%%%%%%%%%%%%%%%%%%%%%%%%%%%%%%%%%%%%%%%%%%%%%%%%%
%\begin{align}
\begin{eqnarray}
\frac{dS_{A}(t)}{dt} &=&  4 \left |B  \right | S_{A} S_{B}~,
\label{3} 
\end{eqnarray}
%%%%%%%%%%%%%%%%%%%%%%%%%%%%%%%%%%%%%%%%%%%%%%%%%%%
\begin{eqnarray}
\frac{dS_{B}(t)}{dt}  &=& -2 S_{n}  \left |A  \right | - \left |B  \right | + 4 \left |B  \right | S_{B}  S_{B}~,
\label{4} 
\end{eqnarray}
%%%%%%%%%%%%%%%%%%%%%%%%%%%%%%%%%%%%%%%%%%%%%%%%%%%
\begin{eqnarray}
\frac{dS_{n}(t)}{dt}  &=& 2 S_{B}\left |A  \right |  + 4 \left |B  \right | S_{B}  S_{n}~. 
\label{5} 
\end{eqnarray}
%\end{align}
%%%%%%%%%%%%%%%%%%%%%%%%%%%%%%%%%%%%%%%%%%%%%%%%%%%
In absence of the non-Hermitian term ($\vec{B}=0$), $\vec{S}. \hat{A}=S_{A}$  is a constant of motion as expected however as $\vec{B}$ becomes finite this is no more true. But, if $S_{A}$ is set to zero initially then it stays zero as a function of time owing to the fact that solution to (\ref{3}) takes the form $S_A(t)= S_A(t_0) \exp [4 \left |B  \right | \int_{t_0}^{t} S_B(t) ]$. Geometrically this fact has two implications: {\cal (i)} all solutions with boundary condition $S_{A}=0$ lies in  the plane spanned by ${\{} \hat{B}, \hat{n} {\}}$,  {\cal (ii)} all such solutions corresponding to pure states will trace out geodesics paths on the Bloch sphere. Lastly, note that all these three equation are invariant under the transformation, $S_{A}\rightarrow -S_{A}$.  This implies that all closed looped trajectories laying on the Bloch sphere which are solutions of this equation, trajectories generated by geometric reflection of these  trajectories about the ${\{} \hat{B}$-$\hat{n} {\}}$-plane are also valid solutions to the same equation. Now it is straightforward to obtain analytic solution for $S_{B}(t)$ and $S_{n}(t)$ for the case of $S_{A}=0$ given by :
%%%%%%%%%%%%%%%%%%%%%%%%%%%%%%%%%%%%%%%%%%%%%%%%%%%
\begin{equation}
S_{B}=   \frac{1}{2} ~\sqrt{\frac{(A^2-B^2)\sin^2 2 \sqrt{A^2-B^2} \ t}{(A-B \cos 2 \sqrt{A^2-B^2} \ t)^2}}~,
\label{6} 
\end{equation}
%%%%%%%%%%%%%%%%%%%%%%%%%%%%%%%%%%%%%%%%%%%%%%%%%%%
\begin{equation}
 S_{n} = \frac{1}{2} \frac{B-A \cos 2 \sqrt{A^2-B^2} \ t}{A- B \cos 2 \sqrt{A^2-B^2} \ t}~.
 \label{7} 
\end{equation}
%%%%%%%%%%%%%%%%%%%%%%%%%%%%%%%%%%%%%%%%%%%%%%%%%%%
It is evident from the form of the solutions that they correspond to non-uniform speed of evolution of the state and being on geodesic path (the great circle) these non-uniformities are expected to take extreme values when compared with any other path on the Bloch sphere. \\
%%%%%%%%%%%%%%%%%%%%%%%%%%%%%%%%%%%%%%%%%%
\subsection{Speed of state evolution in $S_{A}=0$ subspace } %%%%%%%
%%%%%%%%%%%%%%%%%%%%%%%%%%%%%%%%%%%%%%%%%%
 Let us start by considering an explicit non-Hermitian Hamiltonian with real eigenvalue given by $H_{\theta} =  \ \sec(\theta) \ \sigma_{x} +  i  \ \tan(\theta) \ \sigma_{z}$, such that $\hat{A}=\hat{x}$ and $\hat{B}=\hat{z}$,  and we choose an initial state for evolution which respects $S_{A}=0$ given by $\ket{\uparrow}_{y} =\{i/\sqrt{2}, 1/\sqrt{2} \}^{t}$. Since the non-Hermiticity in our case is solely governed by the parameter $\theta$ we need to study optimization of speed of state evolution with respect to this parameter.  To study the quantum speed limit (SQL)~\cite{Mandel1945} we first  start by studying the evolution of geodesic distance between two initially ($t=0$) orthogonal states  (in the $S_{A}=0$ subspace) given by $\delta=  \cos^{-1}(|\braket{\psi(t)|\downarrow}_{y}|)$~\cite{Chru1999} where :

 \begin{equation}
  \ket{\psi(t)}= {N(t)} \ exp[-i H_{\theta} t] \ket{\uparrow}_{y}
  \label{8}
 \end{equation}

and $N(t) =1/ \sqrt{\bra{\uparrow}_{y} exp[i H^{\dagger}_{\theta} t] \ exp[-i H_{\theta}  t] \ket{\uparrow}_{y}}$ is the time dependent normalization derived from eq.(\ref{0}). The distance $\delta$ evaluated as a function of time for different values of $\theta$ given by (see Appendix A):
%%%%%%%%%%%%%%%%%%%%%%
%%%%%%%%%%%%%%%%%%%%%%
\begin{figure}[h!]
  \includegraphics[width=7cm,height=4.5cm]{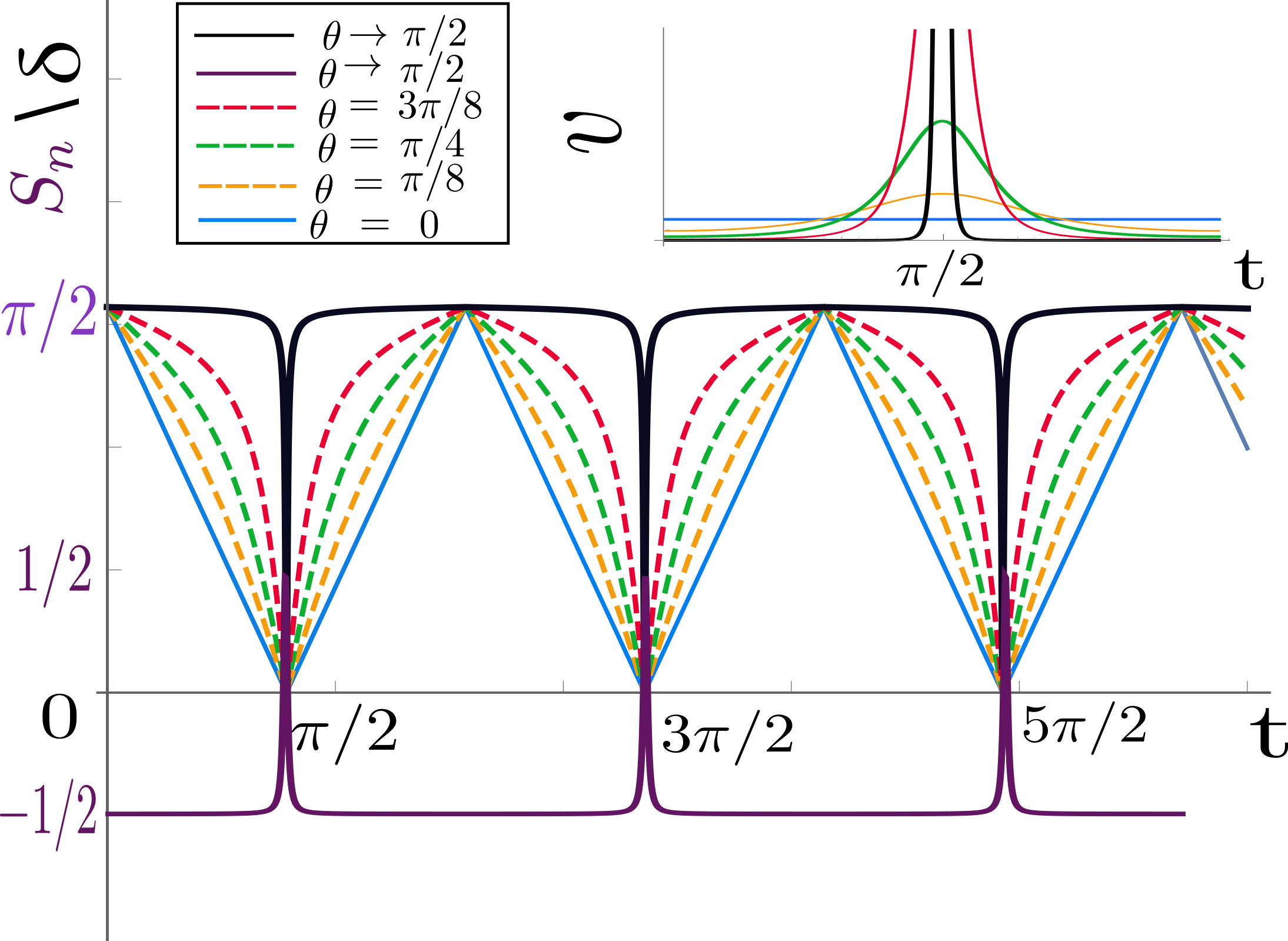}
  \caption{Plot showing variation of $\delta$ as a function of  time for different values of $\theta$. Also, the solution $S_{n}$ in eq.~\ref{6} is plotted as a function of  time showing the evolution of initial state $\ket{\uparrow}_{y} =\{i/\sqrt{2}, 1/\sqrt{2} \}^{t}$ on the Bloch sphere (see Appendix A). Inset shows speed of evolution, $v$, of the initial state $\ket{\uparrow}_{y}$} (in arbitrary units) as a function of time.
  \label{fig:fig1}
\end{figure}
%%%%%%%%%%%%%%%%%%%%%%
%%%%%%%%%%%%%%%%%%%%%%
%%%%%%%%%%%%%%%%%%%%%%%%%%%%%%%%%%%%%%%%%%%%%%%%%%%
\begin{equation}
\delta =\cos^{-1} \left [~{\Big{\vert}} ~ \sqrt{\frac{\sin^{2}(  t) (1- \sin  \theta)}{(1 + \cos(2  t )\sin  \theta)}} ~{\Big{\vert}}~ \right ] .
\label{9}
\end{equation}
%%%%%%%%%%%%%%%%%%%%%%%%%%%%%%%%%%%%%%%%%%%%%%%%%%%
As can be seen from  fig.~\ref{fig:fig1}, the variation of distance, $\delta$ with time is at a constant rate for the Hermitian case ($\theta=0$) while for the case of extreme non-Hermiticity ($\theta\rightarrow \pi/2$) it  mostly shows a vanishing slow variation deep inside the interval $t=[0,\pi/2)$ and $t=(\pi/2, \pi]$ while  depicting a extremely fast  variation corresponding to an almost instantaneous spin-flip (see variation of $S_n$ given by  eq.(\ref{7}) in the plot) process in neighbourhood of  $t=\pi/2$. It is important to note that that physical range of $\theta$ is given by $[0, \pi/2)$ as both $S_{B}$ and $S_{n}$  are ill-defined at $\theta=\pi/2$. As shown by Anandan and Aharonov~\cite{Anandan1990}, the speed of evolution can be determined by expanding $\left |\braket{\psi(t)|\psi(t + \delta t)}\right |^{2} = 1- v^2 \delta t^{2} + O(\delta t^{3})$ ~, where $v = \sqrt{ \  v_{1} + v_{2} + v_{3}}$  is identified as the speed of evolution. Here $v_{1}=(\Delta \vec{A} \cdot \vec \sigma)^{2}$, $v_{2} = (\Delta \vec{B} \cdot \vec \sigma)^{2} $ and $v_{3} = - i \left \langle [\vec{A} \cdot \vec \sigma,\vec{B} \cdot \vec \sigma]_{C} \right \rangle$ (see Appendix B). Hence $v$ for our case is obtained as $v = ({\cos^2 \theta})/({(1+\cos 2 t \ \sin \theta)^2})$.
%%%%%%%%%%%%%%%%%%%%%%
%%%%%%%%%%%%%%%%%%%%%%
  \begin{figure}[h!]
  \includegraphics[width=7cm,height=4.5cm]{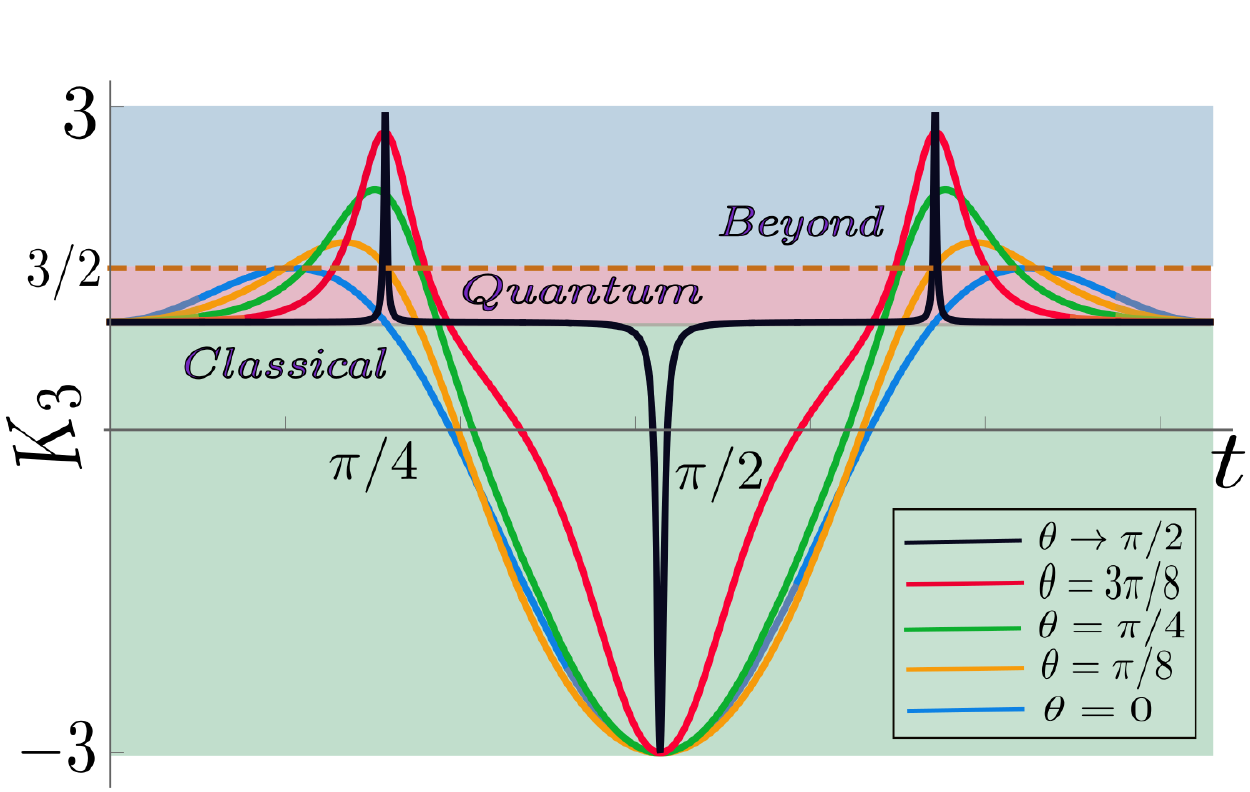}
  \caption{Plot showing variation of LGI parameter as a function of time $t$ for different values of $\theta$,  where the initial state is $\ket{\uparrow}_{y}$ .}
  \label{fig:fig2}
\end{figure}
%%%%%%%%%%%%%%%%%%%%%%
%%%%%%%%%%%%%%%%%%%%%%
Inset of fig.~\ref{fig:fig1} shows the plot of $v$ as a function of time and it depicts the direction correlation of $v $ and $\delta$. It show unbounded acceleration of state evolution in the neighbourhood of $t=\pi/2$ as $\theta \rightarrow \pi/2$. It is important to notice that, unlike the Hermitian case~\cite{Anandan1990,Graefe2012} , the speed of evolution $v$ can not be identified with the standard deviation of energy ($\Delta E$) for non-Hermitian dynamics. Also, note that the third term $v_{3}$ is responsible for extreme acceleration and deceleration of state evolution for $\theta\rightarrow\pi/2$. Hence we have established that $S_{A}=0$ subspace allows for unbounded acceleration of states and we will show now that this fact facilitates an access to the algebraic maxima of LGI. \\
%%%%%%%%%%%%%%%%%%%%%%%%%%%%%%%
\subsection{LGI and its algebraic maxima } %%%%%%
%%%%%%%%%%%%%%%%%%%%%%%%%%%%%%%
 For evaluation of $K_3$ we consider dynamics of TLS with Hamiltonian $H_{\theta}$ and a dichotomic observable for performing the quantum measurement given by $Q=\sigma_{y}$ (anti-parallel to $\hat{n})$. We also assume $t_{1}=0$. We further consider a situation where the temporal spacing between successive measurement is equispaced, i.e., $t_{3}= 2 t_{2} = 2 t$. This leads to $C_{13}= ({\cos 4 t + \sin \theta})/({1+ \cos 4 t \ \sin \theta})$ (see Appendix C).  Note that $C_{13}$ becomes $-1$ at $t = \pi/4$. The LGI parameter at $t = \pi/4$ takes a simple form given by $K_{3}  = 1 + \sin \theta + \sin^{2} \theta$, with $C_{12} = \sin \theta$ and $C_{23} = \sin^2 \theta$. It is easy to see that $C_{12} \rightarrow 1,~C_{23} \rightarrow 1$ and $K_{3} \rightarrow \ 3$ and as $\theta \rightarrow  \pi/2$.  Note that $\theta \rightarrow  \pi/2$ is exactly the limit in which unbounded acceleration of states was observed in fig.~\ref{fig:fig1}. Hence we have established that it is possible to asymptotically approach the conditions, $C_{12}=C_{23}=1$ and  $C_{13}=-1$ simultaneously, which is the key to extremizing $K_3$ as pointed out in the introduction section of this article. It is very interesting to note from fig.~\ref{fig:fig2} that, as  $\theta \rightarrow  \pi/2$,  $K_{3}$ stays at the boundary of classical and quantum dynamics (i.e. $K_3=1$) most of the time and then it momentarily shoots up to extreme values of value of $K_3=-3$ (deep in classical region) or $K_3=3$ ( much beyond the quantum limit). 
%%%%%%%%%%%%%%%%%%%%%%%%
%%%%%%%%%%%%%%%%%%%%%%%%
\begin{figure}[h!]
 \includegraphics[width=7cm,height=4.5cm]{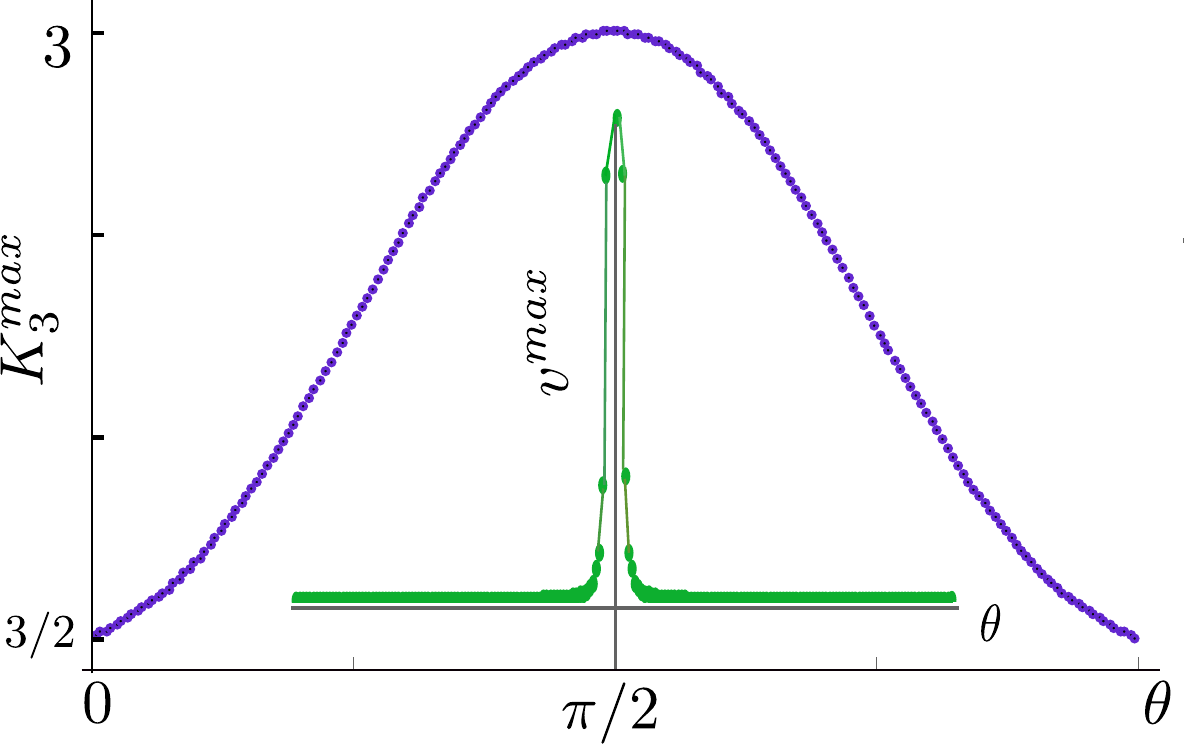}
 \caption{The main panel shows variation of maximum values of LGI parameter} $K^{max}$ in the full parameter space comprising of the parameters space of all possible  Hamiltonian $\hat{H}_{\theta}$, space of all possible initial state and  and space of all possible measurement operator $Q$, while the inset shows variation of $v^{max}$ in the same space as a function of $\theta$.
\label{fig:fig3}
\end{figure}
%%%%%%%%%%%%%%%%%%%%%%%%
%%%%%%%%%%%%%%%%%%%%%%%%
Hence we have not only show that the non-Hermitian dynamics of TLS allows for violation of LGI up to the algebraic bound but we also established a directly connection of this violation with extreme speed of state evolution over geodesic paths. 

Finally, we perform a numerical search over the full parameter space of $K_3$ which comprises of,  {\cal{(i)}} all possible initial state which lie on the Bloch sphere (amounting to two parameters),  {\cal{(ii)}} all possible dichotomic observable $Q$ (two parameters) and {\cal{(iii)}} all possible values of times given by the variables $t_1, t_2$ and $t_3$ (three parameters).Hence it amounts to a scan for finding maximum possible value of $K_3=K_3^{max}$ in the space of these seven parameters for a given $\theta$. Our findings are depicted in fig.~\ref{fig:fig3} which shows $K_3^{max}$ in the main plot. We also perform an independent numerical scan for identifying the maximum velocity $v=v^{max}$ in this parameters space for a given $\theta$ and plot it in the inset. This numerical exercise shows a direct correlation between the maximum violation of LGI and the maximum velocity of state evolution hence reinforcing our analytic finding for the subspace of $S_A=0$ and establishing it as a generic fact. The study also demonstrates that the choice of non-Hermitian Hamiltonian (via choice of $\theta$) single-handedly limits the maximum allowed violation of LGI where the lowest value for the $K_3^{max}$ corresponds to the Luder's bound of $3/2$ corresponding to the case of $\theta=0,\pi$, i.e., the case of Hermitian Hamiltonian and the highest value of $K_3^{max}$ corresponds to $\theta\rightarrow\pi/2$. Hence, an increasing $\theta$ starting from  $\theta=0$ to $\theta=\pi/2$ defines Hamiltonian with an increasing degree of non-Hermiticity defined in the sense of increasing degree of violation of LGI.\\
%%%%%%%%%%%%%%%%%%%%%%

\noindent

%%%%%%%%%%%%%%%%%%%%%%
\section{Embedding of TLS into larger Hilbert space } %%%%%%
%%%%%%%%%%%%%%%%%%%%%%
It is also worthwhile to remark that non-Hermitian Hamiltonian can be simulated using Hermitian systems. Such non-Hermitian Hamiltonian dynamics can be reinterpreted as a dynamics of a subsystem of a higher-dimensional Hermitian system as shown in Ref.~\onlinecite{Ueda,Gunther} (for real eigenvalues of non-Hermitian Hamiltonian) and  Ref. \onlinecite{Ray} (for imaginary eigenvalues). Following these references we consider embedding the TLS into a larger four dimensional Hilbert space $\mathcal{H}_{4}=\mathcal{H}_{2} \otimes \mathcal{H}_{TLS}$ by adding an ancilla which is represented by a Hermitian two level system with  $\mathcal{H}_{2}$ as its space of states. In order to generate the effective non-Hermitian dynamics of the TLS given by $H_{\theta} =  \ \sec(\theta) \ \sigma_{x} +  i  \ \tan(\theta) \ \sigma_{z}$, the  corresponding embedding Hermitian Hamiltonian (ancilla $\otimes$ TLS) reads as :

 \begin{equation}
      H_{T} = \mathbb{I}_{2} \otimes H_{s} + \sigma_{y} \otimes V  
      \label{10}
 \end{equation}
where $H_{s} = \cos \theta  \ \sigma_{x}$ and $V = - \sin \theta \ \sigma_{z} $.  

The idea is to study time evolution generated by $H_{T}$ for those states in  $\mathcal{H}_{4}$ which can be expressed in a form given by  $\ket{\Psi_{T}} = N_{T}\ ( \ \ket{\uparrow}_{z} \otimes \ket{\psi} \,+\, \ket{\downarrow}_{z} \otimes  \eta \ket{\psi} \ )$ where $N_{T}$ is the normalization constant and $\eta$ is such that $\eta \ H_{\theta} = H_{\theta}^{\dagger} \ \eta$. The remarkable fact about this construction is that the action of time evolution operator $exp[-i H_{T} t]$ on the state  $\ket{\Psi_{T}}$ keeps it form invariant such that,
 \begin{equation}
 \begin{split}
\ket{\Psi_{T}(t)} &=  N_{T}\ ( \ \ket{\uparrow}_{z} \otimes exp[-i H_{\theta} \ t] \ket{\psi} ~+ \\
                         &~~~~~~~~~~~  \ket{\downarrow}_{z} \otimes  \eta \ exp[-i H_{\theta} \ t] \ket{\psi} \ ).
\end{split} 
\label{11}
\end{equation}
Hence the state of the isolated TLS evolving under the influence of the  non-Hermitian Hamiltonian $H_{\theta}$ at a time instant $t$ can be extracted by projecting $\ket{\Psi_{T}(t)}$ with operator $ P_{\uparrow} =\ket{\uparrow}_{z} \bra{\uparrow}_{z} \otimes \mathbb{I}_{2}$ and post-selecting over $ \ket{\uparrow}_{z} $.

%This interaction between TLS and ancilla allows to evolve the total initial state $\ket{\Psi_{T}(0)}$ in form invariant manner, and time evolved state then can be written as: $\ket{\Psi_{T}(t)} = N_{T}\ ( \ \ket{\uparrow}_{z} \otimes exp[-i H_{{\color{blue}\theta}} \ t] \ket{\psi} + \ket{\downarrow}_{z} \otimes  \eta \ exp[-i H_{{\color{blue}\theta}} \ t] \ket{\psi} \ )$. 

Since we are interested in those states of TLS which belong to the $S_{A}=0$ subspace so that they could lead to maximal violation of LGI, we consider the initial state ($t=0$) of the TLS to be  $\ket{\psi} =\ket{\uparrow}_{y}$ which is consistent with the discussion above in eq.(\ref{8}) which implies that the  normalization $N_{T} = 1/\sqrt{\braket{\Psi_{T}(0)|\Psi_{T}(0)}} = \cos \theta /(\sqrt{2 (1- \sin \theta)})$ where $\eta = \sec \theta  \ \mathbb{I}_{2} + \tan \theta \ \sigma_{y}$. As mentioned earlier, the state of the TLS at time instant $t$ can be extracted by projecting $\ket{\Psi_{T}(t)}$ with operator $ P_{\uparrow} =\ket{\uparrow}_{z} \bra{\uparrow}_{z} \otimes \mathbb{I}_{2}$ at that instant of time and then post-selecting over $ \ket{\uparrow}_{z} $.\\

The post selection results in a state which describes the sub-ensemble of out comes whose measurement over ancilla resulted in $ \ket{\uparrow}_{z} $. For obtaining a correct description of the non-Hermitian evolution of the state we need to normalize again over this sub-ensemble. The state after the post-selection  is given by  $\ket{\psi_{T}^{\uparrow}(t)} =  N_{T}\ ( \  \ket{\uparrow}_{z} \otimes   \ exp[- i H_{\theta}  t]  \ket{\uparrow}_{y} \ ) / \sqrt{\bra{\Psi_{T}(t)} P_{\uparrow} \ket{\Psi_{T}(t)}}$ which can be rewritten as  $\ket{\uparrow}_{z} \otimes \{ N(t) \ exp[-i H_{\theta}  t] \ket{\uparrow}_{y} \} $ where we have used that fact that $ \sqrt{\bra{\Psi_{T}(t)} P_{\uparrow} \ket{\Psi_{T}(t)}} =  N_{T} \sqrt{\bra{\uparrow}_{y}  \ exp[i H_{\theta}^{\dagger}  t]\ exp[ -i H_{\theta}  t] \ket{\uparrow}_{y}}= N_{T} / N(t)$. Note that $N(t)$ is the same time dependent normalization that appears in eq. (\ref{8}) hence indicating that the fact that the embedding is reliable and the effective non-Hermitian evolution of the state is given by $\ket{\psi(t)}= N(t) \ exp[-i H_{\theta}  t] \  \ket{\uparrow}_{y}$. Now one has to perform further measurement of the form $\mathbb{Q}= \mathbb{I}_{2}\otimes Q$ (where $Q$ represents dichotomic observable for the TLS) on the sub-ensemble  to determine the $K_3$ corresponding to effective non-Hermitian dynamics.\\

For understanding the $\theta\rightarrow \pi/2$ limit (equivalently the $K_{3}\rightarrow 3$ limit) from the perspective of embedding, we rewrite the embedding
Hamiltonian $H_{T} = \mathbb{I}_{2} \otimes \sin \delta \  \sigma_{x}- \sigma_{y} \otimes  \cos \delta \ \sigma_{z} $ and  total  initial state $\ket{\Psi_{T}(0)} = (\sqrt{(1+ \cos \delta)/2}) \ket{\uparrow}_{z} \otimes \ket{\uparrow}_{y} + (\sqrt{(1- \cos \delta)/2}) \ket{\downarrow}_{z} \otimes \ket{\uparrow}_{y}$ in terms of  $\delta$ defined as $\theta = \pi/2 - \delta$. Hence we note that the embedding Hamiltonian remains well defined and corresponding initial state $\ket{\Psi_{T}(0)}$ retains its desired form (representing finite entanglement between the ancilla and the TLS) in the limit of small $\delta$ hence assuring that the proposed embedding of the TLS into a four level system is good enough to asymptotically approach the algebraic maximum of $K_{3}^{max}=3$. We note that in the $\delta\rightarrow 0$ limit, the value of $K_{3}$ to leading order can be shown to be $K_{3}(\delta \rightarrow 0) = 3(1- \delta^{2}/2)$. It should be further noted that at the $\delta = 0 $ point, $\ket{\Psi_{T}(0)}$ becomes a separable state $\ket{\Psi_{T}(0)} =  \ket{\uparrow}_{z} \otimes \ket{\uparrow}_{y}$ and the strategy of generating non-Hermitian evolution by employing a higher embedding itself breaks down.  Hence, only  asymptotically close values to algebraic maximum  of $K_{3}$ can be achieved using this strategy.

\begin{figure}[h!]
 \includegraphics[width=7cm,height=4.5cm]{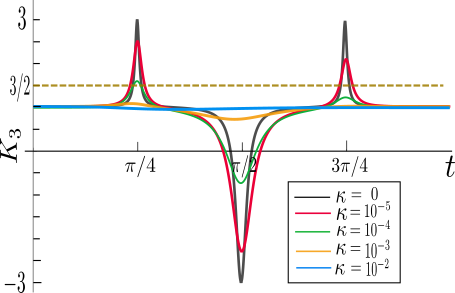}
 \caption{Variation of LGI parameter $K_{3}$ as a function of time for different values of noise parameter $\kappa$ in the extreme non-Hermitian limit (parameter $\theta\rightarrow \pi/2$).}
\label{fig:fig4}
\end{figure}

\section{\label{sec:level4 } Possible influence of energy fluctuation on LGI}

\begin{figure}[h!]
 \includegraphics[width=7cm,height=4.5cm]{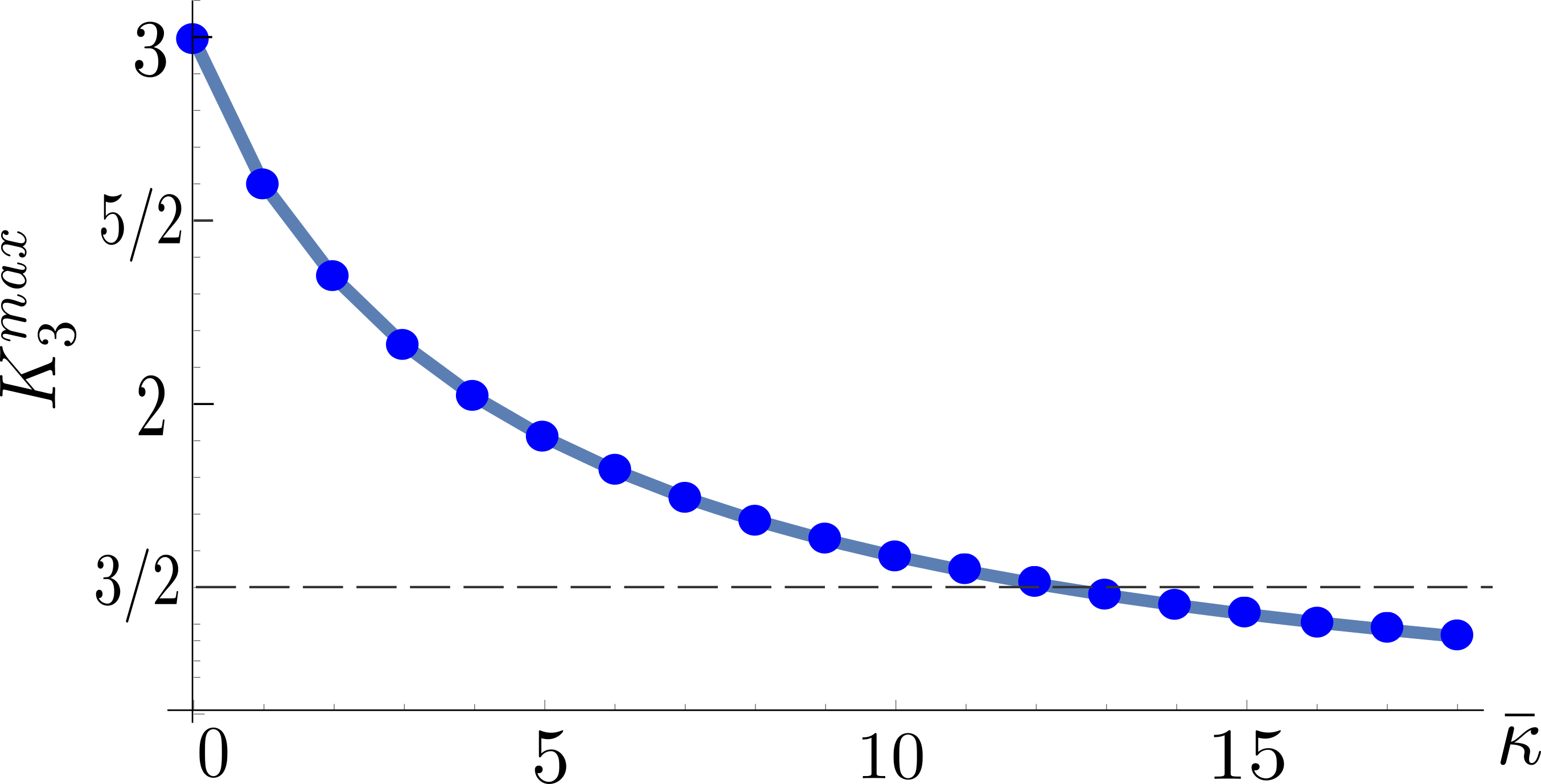}
 \caption{Plot showing the variation of the maximum value of $K_{3}^{max}$ as a function of noise parameter $\bar{\kappa} = \kappa \times 10^{5}$.}
\label{fig:fig5}
\end{figure}

The dynamics considered in the paper corresponds to balanced loss-gain of energy owing to the reality of eigenvalues of the non-Hermitian Hamiltonian. On the other hand, any experimental implementation of temporally resolved balanced loss-gain of energy of a quantum state will be impossible to maintain and it will inevitably suffer from finite temporal fluctuation of energy. To take such effect into account we consider the approach outlined in \cite{Graefe2012}, where eq.(\ref{0}) has been considered with Gaussian white noise. This leads to  modification of evolution equation for the density matrix and it takes the form given by :

\begin{equation}
 \begin{split}
\frac{d \rho_{t}}{dt} &= -i \left [ \vec{A}\cdot \vec \sigma,\rho_{t} \right ] -\left \{ \vec{B}\cdot \vec \sigma,\rho_{t}  \right \} + \\
                         &~~~~~~~~~~~   2 \ tr( \rho_{t}  \vec{B}\cdot \vec \sigma) \rho_{t} + \kappa (\mathbb{I}- 2 \rho_{t})
\end{split} 
\label{12}
\end{equation}

where $\kappa$ is the strength of the noise. Using this equation we recalculate the LGI parameter $K_{3}$ for different values of strength parameter $\kappa$. We found that the point of extreme violation of Luder's bound re-normalizes itself to smaller values and eventually settles at the $K_{3}=1$ line for larger values of $\kappa$ as can be seen from fig.~\ref{fig:fig4}. We also plot the variation of maximum value achievable for the LGI parameter $K_{3}^{max}$ with the scaled noise parameter $\bar{\kappa}= \kappa \times 10^{5}$ (fig.~\ref{fig:fig5}) which provides the value of $ \kappa$ at which the $K_{3}^{max}$ attains the value of 1.5 and eventually saturates to unity. 

\section{Conclusion \& outlook}
To conclude, we have demonstrated that the non-Hermitian dynamics of TLS leads to violation of the Luder's bound of $K_3=3/2$ for LGI.  We further show that the extent of violation can be optimized to asymptotically approach the algebraic maximum of $K_{3}^{max}=3$, which is otherwise only feasible when the Hilbert space is infinite dimensional in the Hermitian case. We established that this extreme violation of LGI is a consequence of unbounded growth of quantum speed of state evolution of TLS owing to the non-Hermiticity  induced non-linear terms in the Bloch equation. Our finding uncovers a completely new view of extreme temporal correlations of two level systems in terms of speed of state evolution. \\

\begin{figure}[h!]
 \includegraphics[width=7cm,height=4.5cm]{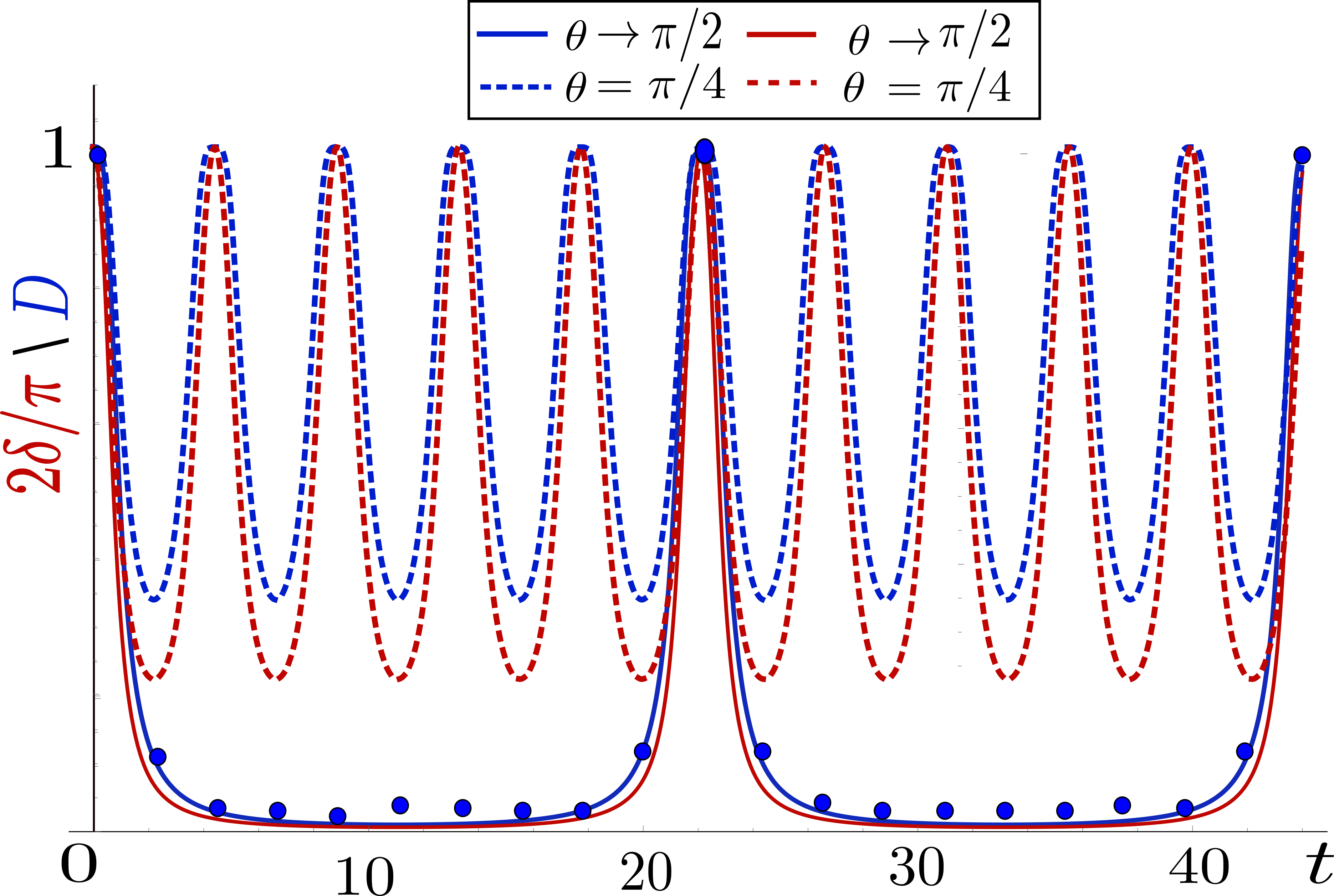}
 \caption{Comparison of geodesic distance $\delta$ and trace distance $D$ (mentioned in Ref \cite{Peng}) plotted as a function of time for different values of parameter $\theta$. Dots (blue) representing experimental data points are extracted from fig.~2(c) of Ref.~\cite{Peng}, fits well with the distance $\delta$ in the limit $\theta \rightarrow \pi/2$.}
\label{fig:fig6}
\end{figure}

We have also shown that such dynamics can be simulated in a realistic quantum system by embedding the non-Hermitian TLS dynamics into higher dimensional Hilbert spaces where states follow unitary time evolution.  We have established that a four dimensional embedding  of the TLS is enough to host $K_{3} \rightarrow 3$ limit which is an encouraging result from the perspective of possible experimental realization of proposed non-Hermitian dynamics. It is worth noting that non-Hermitian Hamiltonians for quantum systems have already been realized experimentally \cite{Tang,Li}. More importantly the non-Hermitian dynamics dictated by eqn. (\ref{0}) has also been implemented experimentally using a photon interferometric network in Ref.~\cite{Peng} in two different ways. Firstly,  they have used the experimental setup for simulating a non-unitary dynamics of a single-qubit, i.e., a non-Hermitian two level system. In a different experimental setup, unitary dynamics of a two-qubit system is realized  and an effective non-Hermitian dynamics of a single qubit is studied via the embedding approach following the proposal of Ref.~[\onlinecite{Ueda,Gunther}]. In order to compare the experiment findings of Ref.~\cite{Peng} and our theoretical work, we compare the plot of the quantity called "distinguishability" defined in their paper which is directly related to the distance $\delta$ defined in eq.(\ref{9}) in an appropriate limit as discussed below. For a fair comparison with the Hamiltonian in Ref.~[\onlinecite{Peng}] we re-scale our Hamiltonian $H_{\theta}\rightarrow \cos\theta\, H_{\theta}$ and then calculate the geodesic distance $\delta$~\cite{Chru1999} of two orthogonal states (namely $\ket{H}=  \{1,0\}^{t}$ and $\ket{V}=  \{0,1\}^{t}$) after evolving them for time  "t". It is worth noting that the distance measure defined in Ref.~\cite{Peng} given by $D[\rho_1,\rho_2]=(1/2) \, Tr[| \rho_1-\rho_2 |]$ is different then the geodesic distance $\delta$ but in ultra non-Hermitian limit ($\theta \rightarrow \pi/2$) they turn out to be the same (see fig.~\ref{fig:fig6}). We extract the experimental data points from fig.~2(c) of Ref.~\cite{Peng} and plot it together with the corresponding value of  $\delta$ evaluated from the experimental inputs of Ref.~[\onlinecite{Peng}] in fig.~\ref{fig:fig6}. We observe an excellent fit of the data with both $\delta$ and  $D$ in the $\theta \rightarrow \pi/2$. This is indeed very encouraging as it not only implies that the experimental realization of this kind of non-Hermitian dynamics is possible, but it also implies that the extreme non-Hermitian limit of $\theta \rightarrow \pi/2$ identified via the extreme limits of $K_3$ could be an experimentally achievable limit. For completeness we also plot  $\delta$ and  $D$ as a function of time in the same plot for value of $\theta$ different then the $\theta \rightarrow \pi/2$ limit which demonstrate that they are not the same quantity in general. 

 Finally we would like to point out that the implication of our findings are of importance for the quantum information and quantum computation community at large which can be understood as follows:\\
 (i) A non-hermitian TLS demonstrates extreme speed up of time evolution of state with respect to its hermitian counterpart,  \\
 (ii) the non-hermitian  dynamics of such a  TLS can be  simulated by appropriately embedding the TLS  into a higher dimensional Hilbert space with Hermitian Hamiltonian. \\
 Hence (i) and (ii) together implies, if we want to speed up the time evolution of quantum state of TLS beyond the limit defined by its dynamics in isolation, we can achieve this by appropriately coupling the TLS to an external degrees of freedom following the prescription given in Eq.\ref{10}. This observation implies  that  the  coupling of  a qubit  to an appropriately chosen external quantum degrees of freedom can be used as a resource for speeding  up of gate operations on the qubit and hence can be of great relevance to fast quantum information processing. \\
%%%%%%%%%%%%%%%%%%%%%%

{\underline{\textbf{Note added :-}}} %%%%%%
%%%%%%%%%%%%%%%%%%%%%%
It should be noted that these results were first reported in an abstract for a talk submitted on June 5, 2019  by one of the authors (S. Das) in the second annual conference on Quantum Condensed Matter held at the Indian Institute of Sciences, Bengaluru (see {http://www.qmatiisc2019.in} and the abstracts therein~\cite{conf}). During the finishing stages of this manuscript, we noted that another work appeared on the arXiv~\cite{Usha2019} which deals with similar idea however our treatment provides a deeper insight into the phenomenon in terms of a renormalized Bloch equation and the quantum speed limit. Also our work presents a Hermitian embedding of the TLS which was not discussed in arXiv~\cite{Usha2019}. \\
%%%%%%%%%%%%%%%%%%%%%%

\section*{Acknowledgements}
 %%%%%%
%%%%%%%%%%%%%%%%%%%%%%
It is a pleasure to thank Poonam Mehta for several discussions and comments. AVV would like to thank the Council of Scientific and Industrial Research (CSIR), Govt. of India for financial support and S.D. acknowledges the kind hospitality of the Condensed Matter and Statistical Physics Division at the Abdus Salam ICTP under the Associateship Scheme during the finishing stages of this work. S.D. would like to acknowledge support from DST-SERB MATRICS grant (MTR/2019/001043) and ARF grant from IISER Kolkata.

%%%%%%%%%%%%%%%%%%%%%%%%%%%%%%%%%%%%%%

\appendix
\section{The geodesic distance}

%In order to define the geodesic distance we first notice that evolution of state $\ket{\uparrow} =\{i/\sqrt{2}, 1/\sqrt{2} \}^{T}$, owing to the Hamiltonian 
%$H =  i  \ \tan(\theta) \ \sigma_{z} + \ \sec(\theta) \ \sigma_{x} $. As shown in FIG. \ref{fig:fig4} the state $\ket{\uparrow} =\{i/\sqrt{2}, 1/\sqrt{2} \}^{T}$ always 
%stays on big circle laying in $y-z$ plane under evolution. Now we can define the 

Geodesic distance between two arbitrary pure states is defined as:  $\delta=  \cos^{-1}(|\braket{\psi|\phi}|)$~\cite{Chru1999}. The two states here are: $ \ket{\psi (t)}= N(t)\exp[i H_{\theta} t] \ket{\uparrow}_{y}$ and $\ket{\downarrow}_{y} =\{-i/\sqrt{2}, 1/\sqrt{2} \}^{T}$. The state $\ket{\psi (t)}$  is given by 
\noindent
\begin{eqnarray}
\ket{\psi(t)} &=& N(t) \begin{Bmatrix}
 i (\cos t-\sin t \ (\cos \theta -(1-\sin \theta) \tan \theta)) \\ \cos t+\sin t (\sec \theta - \tan \theta)
\end{Bmatrix} \nonumber
\end{eqnarray}
where $N(t) =\sqrt{\frac{1+\sin \theta}{2+2 \cos 2t \  \sin \theta}}$. The time evolution of this state is shown in fig.~\ref{fig:fig4}. Also, the state $\ket{\psi} = \{-i/\sqrt{2}, 1/\sqrt{2} \}^{T}$ for $t = \pi/2$, irrespective of the values of $\theta$. Hence the distance $\delta$ reads as 
\begin{eqnarray}
\delta &=& \cos^{-1} \left [~{\Big{\vert}} ~ \sqrt{\dfrac{\sin^{2}( \ t) (1- \sin \ \theta)}{(1 + \cos(2 \  \ t )\sin \ \theta)}} ~{\Big{\vert}}~ \right ]~. \nonumber
\end{eqnarray}

\begin{figure}[h!]
\includegraphics[width=7cm,height=5cm]{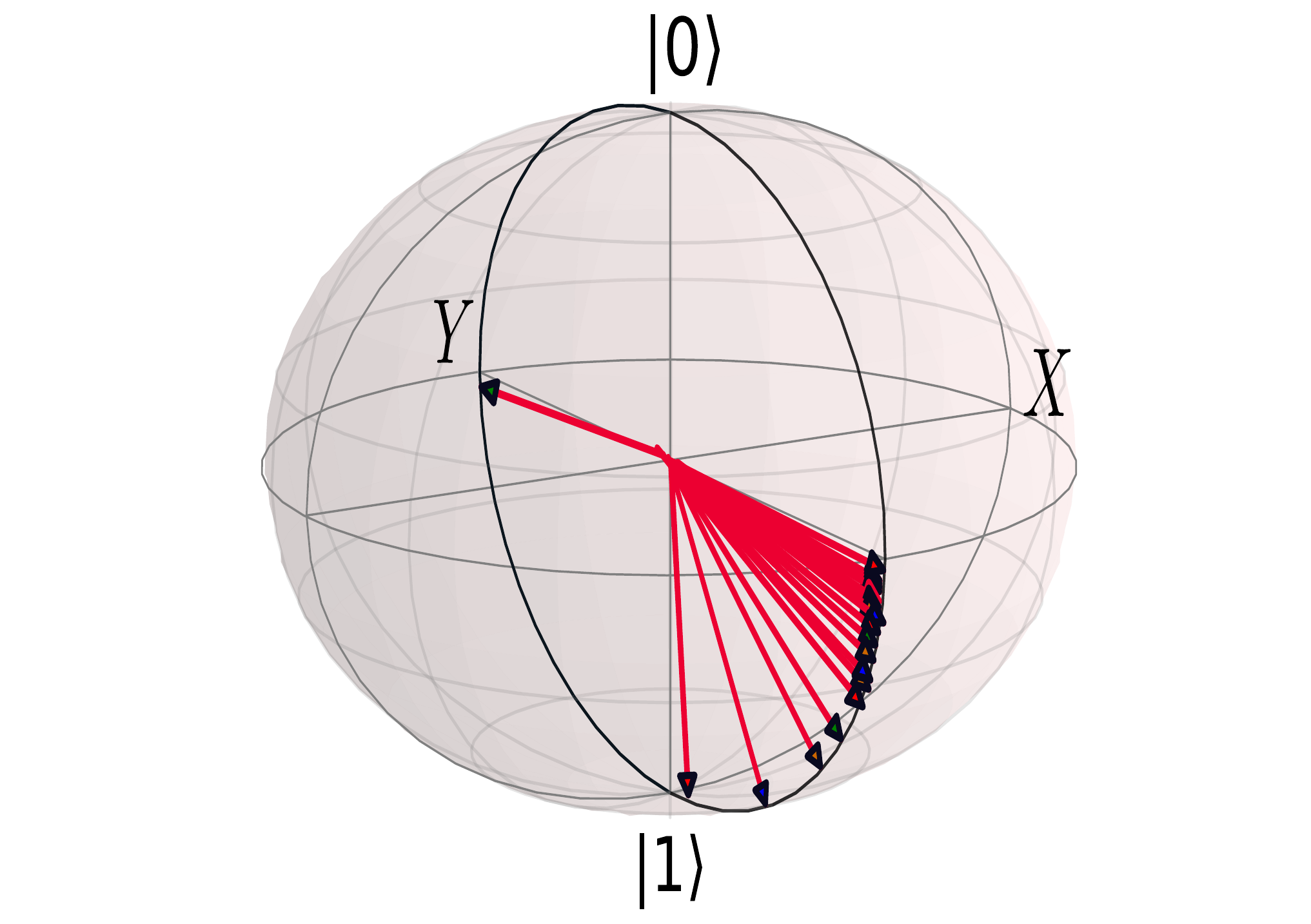}
 \caption{ Evolution of state $\ket{\uparrow}_{y} =\{i/\sqrt{2}, 1/\sqrt{2} \}^{T}$, owing to the Hamiltonian $H_{\theta} =  i  \ \tan(\theta) \ \sigma_{z} + \ \sec(\theta) \ \sigma_{x} $ following the great circle in $y-z$ plane. State at a given time t is represented as red arrows with  black arrowheads. These arrows have been plotted for equal time difference and for the time steps of the order of $10^{-2}$ for a total evolution time of $t= \pi/2$. It shows the sudden change of of the state in the last time step hence demonstrating the accelerated motion.}
 \label{fig:fig7}
\end{figure}

\section{The speed of evolution}

For evaluating the speed of evolution we need to calculate the overlap between  state a time $t$ give by $\ket{\psi_{t}}$ and state at time $t+\delta t$ given by $\ket{\psi_{\delta t}}$  which reads as
\begin{eqnarray}
p(\delta t) &=&  \left |\braket{\Psi_{t}|  \Psi_{t+\delta t } }\right |^{2} =\frac{ \bra{\Psi_{0}}U \ket{\Psi_{0}} \bra{\Psi_{0}}U^{\dagger} \ket{\Psi_{0}}}{\braket{\Psi_{0}|\Psi_{0}} \bra{\Psi_{0}}U^{\dagger}U\ket{\Psi_{0}}}  \nonumber \end{eqnarray} 
where $U = e^{-i H_{\theta} (t+ \delta t)} = \cos (t+\delta t) \ \mathbb{I} - i \ H_{\theta} \ \sin (t+ \delta t) $ since $H_{\theta}^{2} =H_{ \theta}^{\dagger \ 2} = \mathbb{I}$ for our case. Here we have considered normalized state, hence $\braket{\Psi_{t}|\Psi_{t}}= 1$. Now, we can expand the denominator as well as numerator in $\delta t$. Then writing the whole expression upto second order in $\delta t$ gives us three terms. Now, comparing with the expansion discussed above and  $\left |\bra{\psi(t)}  \ket{\psi(t + \delta t)}\right |^{2} = 1- v^2 \delta t^{2} + O(\delta t^{3})$, one identifies $v$ as the speed of the evolution of the state. For the given Hamiltonian and initial state (discussed in the main text), we arrive the following expression for  the speed of evolution given by  
\begin{eqnarray}
v = \dfrac{\cos^2 \theta}{(1+\cos 2 t \ \sin \theta)^2}~.\nonumber
\end{eqnarray}

%\vskip 1cm
\section{Correlation optimization}

%\section{Appendix C: Correlation Optimization}

The two time correlation $C_{ij}$ is given by $\sum_{Q(t_i),Q(t_j)=\pm 1} P_{ij}(Q(t_i),Q(t_j))$, where the joint probability is defined as 
% $P_{ij}(Q(t_i),Q(t_j)) =P_{ij}(+/-,+/-) =$ \\ 
 \begin{eqnarray}  P_{ij}(Q(t_i),Q(t_j)) &=& P_{ij}(+/-,+/-) \nonumber \\
 && \!\!\!\!\!\!\!\! \!\!\!\!\!\!\! \!\!\!\!\!\!\! \!\!\!\!\!\!\! \!\!\!\!\!\!\! \!\!\!\!\!\!\! \!\!\!\!\!\!\!= \dfrac{|\bra{+/-}e^{-iH_{\theta}t_{}}\ket{+/-}|^2|\bra{+/-}e^{-iH_{\theta}t_i}\ket{\psi_{I}}|^2}{ \bra{\psi_{I}} e^{iH_{\theta}^{\dagger}t_i}e^{-iH_{\theta}t_i}\ket{\psi_{I}} \bra{+/-}e^{iH_{\theta}^{\dagger}t_ji}e^{-i H_{\theta} t_{ji}}\ket{+/-}}
~,\nonumber  \end{eqnarray}   \\

 where $\ket{\psi_{I}}$ is the initial state.
Now $C_{ij}$  are evaluated for the case of equal time spacing defined by $t_{3}-t_{2}= t_{2}- t_{1}= t$, with $t_{1}=0$ with inital state taken as the $\ket{\psi_{I}}=\ket{\uparrow}_{y} =\{i/\sqrt{2}, 1/\sqrt{2} \}^{T}$ for the Hamiltonian $H_{\theta}$ and measurement operator $Q=\sigma_y$ given in the main text. The joint probabilities corresponding to the correlation $C_{12}$ are given by
\begin{eqnarray}
 P_{12}(+,+)  &=&  \dfrac{\cos^{2}t (1+\sin \theta)}{1+\cos 2t \sin \theta}~, 
\nonumber\\
 P_{12}(+,-) &=&  -\sin^{2}t (-1 + \sin \theta)/(1 + \cos 2t \  \sin \theta)~,
  \nonumber\\
 P_{12}(-,+) &=& P_{12}(-,-)=0~. \nonumber
  \end{eqnarray}
%and $P_{12}(-,+)= P_{12}(-,-)=0$.  \\
Similarly, joint probabilities for the correlations $C_{23}$ and $C_{13}$ reads as follows 
\begin{eqnarray}
 P_{23}(+,+) &=& \dfrac{\cos^{4}t (1+\sin \theta)^2}{(1+\cos 2t \  \sin \theta)^2}
\nonumber\\
 P_{23}(+,-) &=& -\dfrac{\cos^{2}t \  \sin^{2} t \ (-1+\sin \theta) (1+\sin \theta)}{(1+\cos 2t \  \sin \theta)^2} 
\nonumber\\
 P_{23}(-,+) &=& \dfrac{\sin^{4}t (-1+ \sin \theta) (1+ \sin \theta)}{(-1 + \cos 2t \sin\theta) (1 + \cos 2t \ \sin \theta)} 
\nonumber\\
 P_{23}(-,-) &=& -\dfrac{\cos^{2}t \sin^{2}t (-1+\sin \theta)^2}{(-1+\cos 2 t \  \sin \theta) (1+\cos 2t \  \sin \theta)}
\nonumber\\
 P_{13}(+,+) &=&  \dfrac{\cos^{2} 2t  (1+\sin \theta)}{1+\cos 4t \ \sin \theta}
\nonumber\\
 P_{13}(+,-)&=& -\dfrac{\sin^{2} 2t (-1+\sin \theta)}{1+\cos 4t \ \sin \theta}
\nonumber\\
 P_{13}(-,+)&=& 0 \quad {\textrm{and}} \quad 
  P_{13}(-,-)=0~. \nonumber
 \end{eqnarray}  
 Then the corresponding correlations take the form
 \begin{eqnarray}
C_{12} &=& \dfrac{\cos 2t+ \sin \theta}{1+ \cos 2t \  \sin \theta}~,\nonumber\\
C_{13} &=& \dfrac{\cos 4 t + \sin \theta}{1+ \cos 4 t \  \sin \theta}~, \nonumber
\end{eqnarray}
 and
\begin{widetext}
\begin{eqnarray}
C_{23} &=& -\frac{\cos^{2}2t \ \cos^{2}\theta \  \sin \theta +\sin^{2}2t \ \sin^{2}\theta+ \cos 2t (\cos^{2}\theta +\sin^{2}2t \sin \theta)}{(-1+ \cos2t \ \sin \theta) (1+\cos 2t \sin\theta)^2}~. \nonumber
\end{eqnarray}
\end{widetext}
Now putting $t = \pi/4$, the joint probabilities reduce to
\begin{eqnarray}
 P_{12}(+,+) &=&  \dfrac{1}{2} (1 + \sin \theta) \nonumber\\
 P_{12}(+,-) &=&  \dfrac{1}{2} (1- \sin \theta) \nonumber\\
P_{12}(-,+)&=& P_{12}(-,-)=0 \nonumber
\end{eqnarray}
 Similarly, joint probabilities for the correlations $C_{23}$ and $C_{13}$ reduce to 
\begin{eqnarray}
P_{23}(+,+) &=& \dfrac{1}{4} (1+\sin \theta)^2 \nonumber \\
P_{23}(+,-) &=& \dfrac{\cos^{2} \theta}{4} \nonumber \\
P_{23}(-,+) &=& \dfrac{\cos^{2} \theta}{4} \nonumber \\
P_{23}(-,-) &=&\dfrac{1}{4} (1-\sin \theta)^2 \nonumber \\
P_{13}(+,+)&=& 0 ~ \nonumber\\
 P_{13}(+,-)&=& 1~ \nonumber \\
 P_{13}(-,+)&=& 0 \quad {\textrm{and}} \quad P_{13}(-,-)=0 \nonumber 
\end{eqnarray}
and the correlations are given by  $C_{12} = \sin \theta, C_{23} = \sin^2 \theta$ and $C_{13} =-1$.

%\nocite{*}
%\bibliography{TemporalCorrelations}
%\bibliographystyle{plain}

\end{document}